\documentclass[twocolumn,prb,showpacs,floatfix,aps]{revtex4} %
\usepackage{amsmath,eucal}
\usepackage{graphicx}
\usepackage{pifont}
\usepackage{verbatim}
\usepackage[latin1]{inputenc}
\usepackage{subfigure}
\usepackage{rotating}
\begin{document}

\title{Mechano-switching devices from carbon wire-carbon nanotube junctions}
\author{J. Prasongkit}
\email{jariyanee.prasongkit@npu.ac.th}
\affiliation{Division of Science, Faculty of Liberal Arts and Science, Nakhon Phanom University, Nakhon Phanom, 48000, Thailand}
\affiliation{Condensed Matter Theory Group, Department of Physics and Astronomy,Uppsala University, SE-75120 Uppsala, Sweden}
\author{A. Grigoriev}
\affiliation{Condensed Matter Theory Group, Department of Physics and Astronomy,Uppsala University, SE-75120 Uppsala, Sweden}
\author{R. Ahuja}
\affiliation{Condensed Matter Theory Group, Department of Physics and Astronomy,Uppsala University, SE-75120 Uppsala, Sweden}
\altaffiliation{Applied Material physics, Department of Materials and Engineering, Royal Institute of Technology (KTH), 10044, Stockholm, Sweden}

\date{\today}

\begin{abstract}
Well-known conductive molecular wires, like cumulene or polyyne,  provide a model for interconnecting molecular electronics circuit. In the recent experiment, the appearance of carbon wire bridging two-dimensional electrodes - graphene sheets - was observed [PRL 102, 205501 (2009)], thus demonstrating a mechanical way of producing the cumulene. In this work, we study the structure and conductance properties of the carbon wire suspended between carbon nanotubes (CNTs) of different chiralities (zigzag and armchair), and corresponding conductance variation upon stretching. We find the geometrical structure of the carbon wire bridging CNTs similar to the experimentally observed structures in the carbon wire obtained between graphene electrodes. We show a capability to modulate the conductance by changing bridging sites between the carbon wire and CNTs without breaking the wire. Observed current modulation via cumulene wire stretching/elongation together with CNT junction stability makes it a promising candidate for mechano-switching device for molecular nanoelectronics.
\end{abstract}


\maketitle

\section{Introduction}
A carbon-based materials have now become the most promising candidates for nanoelectronic applications. Applications of carbon nanotubes (CNTs) and graphenes have already emerged; for example, field effect transistors (FETs) \cite{Javey2003,Misewich2003,Ouyang2007}, electrical interconnects \cite{Avouris2007}, and sensors \cite{Siwy2010,Prasongkit:2011}. However, linear carbon wires; namely, cumulene and polyyne, received poor attention owing to the difficulty of getting access to the pure wire \cite{Haley:2010jc}. The cumulene, in particular, was proposed as the ideal molecular wire \cite{Lang:1998}. However, it is still challenging to create such a metal-molecule junction in the experiment, becuse cumulene wire is generally unstable.

The appearance of the cumulene bridging CNTs or graphene was recently observed in the experiment \cite{Jin:2009p592,Chuvilin2009,Troiani:2003p990,Marques:2004p465} by using electron irradiation inside a high resolution transmission electron microscope (TEM), thus demonstrating a mechanical way of producing the carbon wire. An axial strain on the graphene or CNT was induced by the high-energy electron beam of TEM resulting in the fracture, and led to the establishing carbon wire-CNT junction \cite{Jin:2009p592,Troiani:2003p990}. The observed carbon wires were more stable than those produced with previous approaches; however, the long wire ($>$10 carbon atoms) appeared to be unstable \cite{Troiani:2003p990}. Recently, the carbon wire bridged between graphene sheets was shown to perform as a bistable switch \cite{Standley2008,Zhang:2012}, operating for many thousands of cycles without degradation, which can be interesting from an application point of view.

The theoretical investigations have been carried out on the structure of carbon wire-CNT junctions \cite{Marques:2004p465,Enyashin:2005ve}. Before breaking of the junction, the carbon wire bridging between CNTs or graphene sheets can transform into either cumulene or polyyne \cite{Ajayan:1998p1241, Jin:2009p592,Marques:2004p465}. By using the non-equilibrium Green's function (NEGF) approach, Khoo \emph{et. al}\cite{Khoo:2008p688} observed negative differential resistance in the carbon wire-capped CNT junction, bonded through sp$^{3}$ bond. Recently, the carbon wire connected to graphene sheets \cite{Zhang:2010fm,Shen2010} (representing infinite radii of CNT) have been theoretically studied, revealing different electronic functions such as molecular switches \cite{Erdogan:2011bt}, molecular rectifiers \cite{Zeng:2011jo}, and molecular spintronics devices\cite{Zanolli2010}. Carbon wires connected to gold \cite{Prasongkit:2010p780,Crljen:2007p452}, lithium \cite{Zhang13} and fullerene \cite{Wang:2011jf} electrodes have also been investigated.

In this paper, we present the transport properties of the short carbon wire suspended between CNT electrodes by employing NEGF technique based on density functional theory. Varying the gap width between the electrodes changes the wire structure and contact geometries in the junction. The latter is specific for CNT electrodes, in which the bridging site can appropriately adjust to the change of junction lengths. We have focused on the variation in conductance of the carbon wire via repeated compression/elongation of the junction, effectively leading to a prominent difference in conductance. We consider here the zigzag (4,0)CNT and armchair (4,4)CNT, due to a very small diameter of the CNTs observed in the experiment before fracture. \cite{Troiani:2003p990}. The influence of CNT chiralities (zigzag and armchair) on conductance has been investigated. In addition, oscillating behavior of conductance, typical for cumulenes of different lengths \cite{Lang:1998,Lang:2000tp,Emberly2009} is reproduced. Upon junction stretching, the current-voltage characteristics of the carbon wire-zigzag junctions show the high- and low-conduction states (referred as ON and OFF states) at elevated bias, corresponding to the cumulene and polyyne structures, respectively. It should be emphasized that we produced the ON and OFF states by changing the wire configuration without breaking wires, while Standley et al.\cite{Standley2008} have demonstrated experimentally that the switch works by breaking the cumulene wire.

Experimental studies of the graphene and CNT edge have revealed several complex rearrangements \cite{Caglar,Huang2009,Liu:2009iu}. The migration of carbon wire along the graphene edge was observed in the experiment \cite{Jin:2009p592}: the carbon wire could jump along the graphene or CNT edges with a change of bonding site at the junction due to a strain accumulating along the chain. We show the conductance of the C$_5$ wire connected to all possible bridge sites on the (4,0)CNT.  
Interesting features in the electron transport properties of the carbon wires are revealed, including the possibility of current modulation via carbon wire shrinking and stretching without breaking wires. We employ a detailed analysis of transport channels in carbon wire-CNT junctions to explain our results.

\section{Computational Methods}

\begin{figure}[tp]
\begin{minipage}[ht]{1.0\linewidth}
\begin{center}
\includegraphics[width = 8 cm]{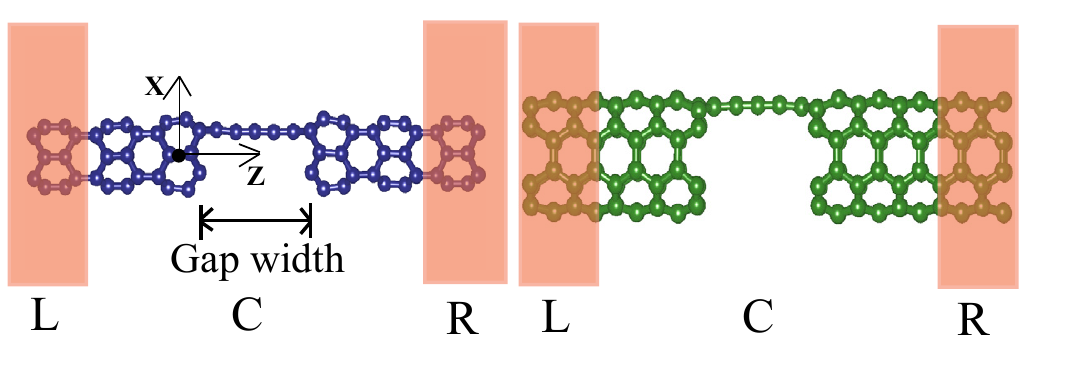}
\end{center}
\end{minipage}
\caption{Two-probe systems for measuring the conductance of carbon wires, connected to semi-infinite zigzag (4,0)CNT (left panel) and armchair (4,4)CNT (right panel).}\label{grap_unit}
\end{figure}

In this section, we briefly describe details of the computational method and the geometrical setup procedure performed in the present work.

The carbon wire-CNT junction is referred as a two-probe system, divided into three regions: the left and right electrodes, and the central region (see Fig. \ref{grap_unit}). To construct the carbon wire-CNT junction, an infinite cumulene and CNT were first optimized separately, and then the short wire was placed between electrodes. The central region includes the CNT on either side of the junction in order to ensure that the perturbation effect from the carbon wire edge is sufficiently screened. The carbon wire-CNT junction was optimized again, allowing all atoms in the central region to relax. Then, the gap width between electrodes was varied with a step of 0.2 \AA.

All optimizations were carried out by using density functional theory (DFT) as implemented in the SIESTA code\cite{Soler2002}. Our calculation was performed within the non spin-polarized generalized gradient approximation (GGA) \cite{Perdew1996} method with a single-$\zeta$ with polarization (SZP) basis, which has already been presented its validity for the short carbon wire-CNT junction \cite{Khoo:2008p688}. The atomic core electrons were modeled with Troullier-Martins norm-conserving pseudo potential \cite{Troullier1991}, and valence states are 2s2p for C. The real-space integrations were performed using 170 Ry cutoff, assuring the energies and forces were converged. All atoms in the central region were allowed to move until the forces were less than 0.01 eV/\AA.

\begin{figure}[tp]
\begin{minipage}[ht]{1.0\linewidth}
\begin{center}
\includegraphics[width = 8.5 cm]{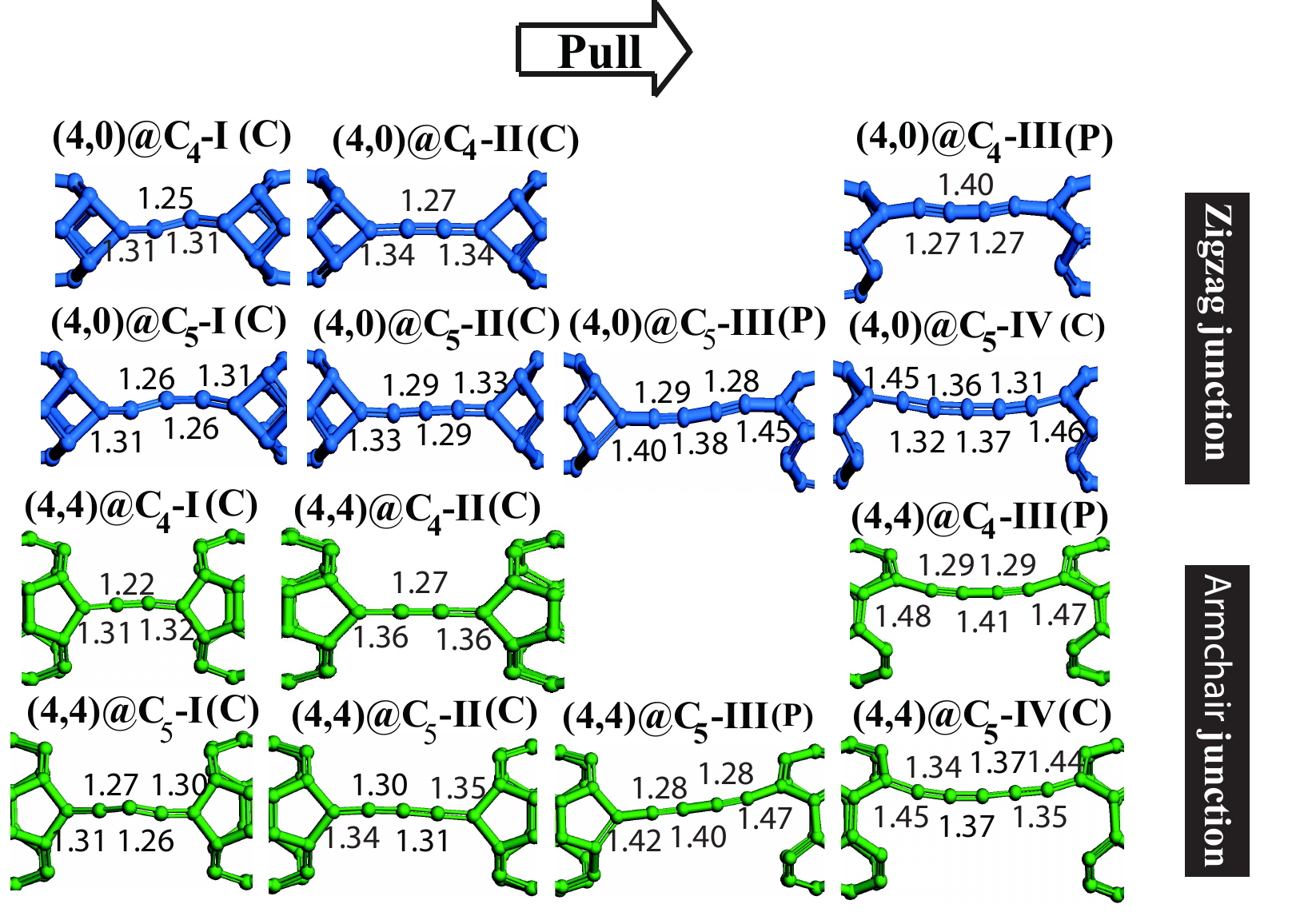}
\end{center}
\end{minipage}
\caption{Various structures of the short carbon wires (C$_{4}$ and C$_{5}$) bridged between zigzag (4,0)CNTs and armchair (4,4)CNTs, in which the gap width is varied. Different geometrical structures and bridging sites of the carbon wires are represented by I, II, III, etc. The linear carbon wire can transform into either cumulene ``C" or polyyne ``P". }\label{sum}
\end{figure}

The transport calculation was performed with the SMEAGOL code \cite{Rocha:2005,Rocha:2006}, based on the combination of the NEGF and DFT. The basis set and the real-space integrations used in the electron transport calculation were the same as that of geometrical optimization part. We have performed test calculations by using a double-$\zeta$ with polarization (DZP) basis set, resulting in only minor change of the transmission function.

The current through the junction was calculated using the Landauer-Buttiker formula \cite{Datta1995,Brandbyge2002}:

\begin{equation}
I=\frac{2e}{h}\int^{\infty}_{-\infty} dE [f(E,\mu_L)-f(E,\mu_R)]T(E,V),
\end{equation}
where $\mu_L$ and $\mu_R$ are the electrochemical potentials of the left and right electrodes, respectively. $T(E,V)$ is the transmission coefficient at energy $E$ and bias voltage $V$, which is evaluated as
\begin{equation}
T(E,V)=\mbox{Tr}[G(E)\Gamma_{L}G^{\dag}(E)\Gamma_{R}],
\end{equation}
where $G(E)$ and $G^{\dag}(E)$ are retarded and advanced Green's function of the central region.

\section{Atomic structures}\label{sec3}

We categorize our systems into two groups: the carbon wire connected to the zigzag and armchair CNTs. We focus on the interval of 1D carbon wire existence. Pulling the wire beyond this interval results in the wire breakage, while compressing the wire more leads into folding structures. Fig.~\ref{sum} demonstrates the various wire structures and contact geometries of the  C$_{4,5}$ wire connected to the zigzag (4,0)CNT and to the armchair (4,4)CNT, varying the gap width. The zigzag and armchair junctions are labeled as (4,0)@C$_{n}$ and (4,4)@C$_{n}$ respectively, where $n$ is the number of carbon atoms in the wire. Different geometrical structures of the wires are labeled by I, II, III, etc. The linear carbon wire can take either cumulene ``C" or polyyne ``P" structures.

Let us first discuss geometrical structure of the carbon wire-zigzag(4,0)CNT junction. As illustrated in Fig.~\ref{sum}, the C$_{4}$ wire is slightly bent due to compression of the junction, labelled as (4,0)@C$_{4}$-I(C), and then becomes the straight-wire when the junction is gradually pulled apart, labelled as (4,0)@C$_{4}$-II(C). These C$_{4}$ wires, forming two bonds to each side of CNTs, are the cumulene structures with double bonds between neighbouring atoms. The central bond length of C$_{4}$ wire can vary in the range of $\sim$ 1.23 \AA \ - 1.36 \AA \, depending on the variation of gap width. Note that the bond length alternation of the C$_4$ wire cannot be observed, since there is only one central bond in the wires. However, we have tested for C$_{6,8}$ wires, showing the bond length alternation of $\sim$ 0.03-0.05 \AA \ upto the gap width. Before wire rupture, the C$_{4}$ wire forms one bond to each side of electrodes, labelled as (4,0)@C$_{4}$-III(P). At this step, the average bond lengths of the C$_{4}$ wires are 1.27 \AA \ and 1.40 \AA, which is a polyyne structure due to the triple and single bond alternation.

For the C$_{5}$ wire, as seen in Fig.~\ref{sum}, the bent wire is observed in the compressed junction, labelled as (4,0)@C$_{5}$-I(C). Then it becomes straight while pulling the junction, labelled as (4,0)@C$_{5}$-II(C). These C$_5$ wires, forming two bonds to each side of CNTs, show the cumulene structure. During compressing and pulling of the (4,0)@C$_{5}$-I(C) and (4,0)@C$_{5}$-II(C), the bond lengths of the cumulene wires vary in the range of $\sim$ 1.26 \AA \ - 1.34 \AA. There is no bond length alternation for the odd cumulene wire.

If we continue to stretch the junction, the C$_{5}$ wire forms one bond to one side of CNTs and two bonds to the other side, labelled as (4,0)@C$_{5}$-III(P). Note that this configuration is not even metastable for the C$_{4}$ wire due to broken symmetry. The average bond lengths of the C$_{5}$  wire are $\sim$ 1.28 \AA \ and 1.42 \AA, revialing the polyyne structure. In particular, the bonding geometry of the (4,0)@C$_{5}$-III(P) is similar to a transition state in the possible migration pathway for the migration of a 3\% strained carbon chain along the graphene sheet with zigzag edge reported in Ref. \citenum{Jin:2009p592}. We would like to emphasize that a curvature of the CNT plays an important role in the structure of the wire at this stage of elongation: the tube edge becomes capped instead of being an open-ended one. This results in a long bond along a chord across the CNT towards the next hexagon, thus defining a single bond to the wire and yielding shorter triple bonds in polyyne. The far more symmetric cumulene structure is less affected by the CNT curvature, but the bonding geometry is changed as compared to graphene, namely the wire is bonded to the site between two adjacent hexagons. Before breaking of the contact, the C$_{5}$ wire forms one bond to each side of CNTs, and becomes the cumulene structure again, (4,0)@C$_{5}$-IV(C). Typically, bond lengths of single, double, and triple bonds are 1.54 \AA, 1.34 \AA \ and 1.2 \AA, respectively \cite{Hino2003}. Before the wire fractures, the central bond lengths of (4,0)@C$_{5}$-IV(C) are stretched so that those bond lengths can be $\sim$ 1.40-1.43 \AA, which are intermediate between single bond and double bond.

In the case of armchair junction, the wire structures and contact geometries at each step of varying the gap width are similar to that of the zigzag junction (Fig.~\ref{sum}); thus, both zigzag and armchair junctions are labeled in the same way. Furthermore, we find that the geometrical structures of the carbon wire-graphene junctions are also similar to those of the carbon wire-CNT junction discussed above.

In conclusion, the gap width variation leads to a change in the wire structures and bonding geometries of the carbon wire connected to the zigzag (4,0)CNTs or armchair (4,4)CNTs. Moreover, the effect of odd-even numbered wires has played an important role for a difference in the wire structure and bonding at the junction for both cases. The observed bond lengths of the wires agree well with previous works \cite{Senapati:2005p2333,Molder:2004dw,Khoo:2008p688}. In particular, the geometrical structure of the carbon wire bridging CNTs are similar to the experimentally observed structure of the cumulenes connected to the graphene, \cite{Jin:2009p592} showing pathways for the migration of the carbon wire along the graphene edge. Differences in the geometry of the junctions as compared to the reported for the graphene are attributed to the finite curvature of the CNT electrodes.
\section{Electron transport properties}
\subsection{Zigzag vs. armchair junction}

\begin{figure*}[tp]
\begin{center}
\includegraphics[scale = 0.75]{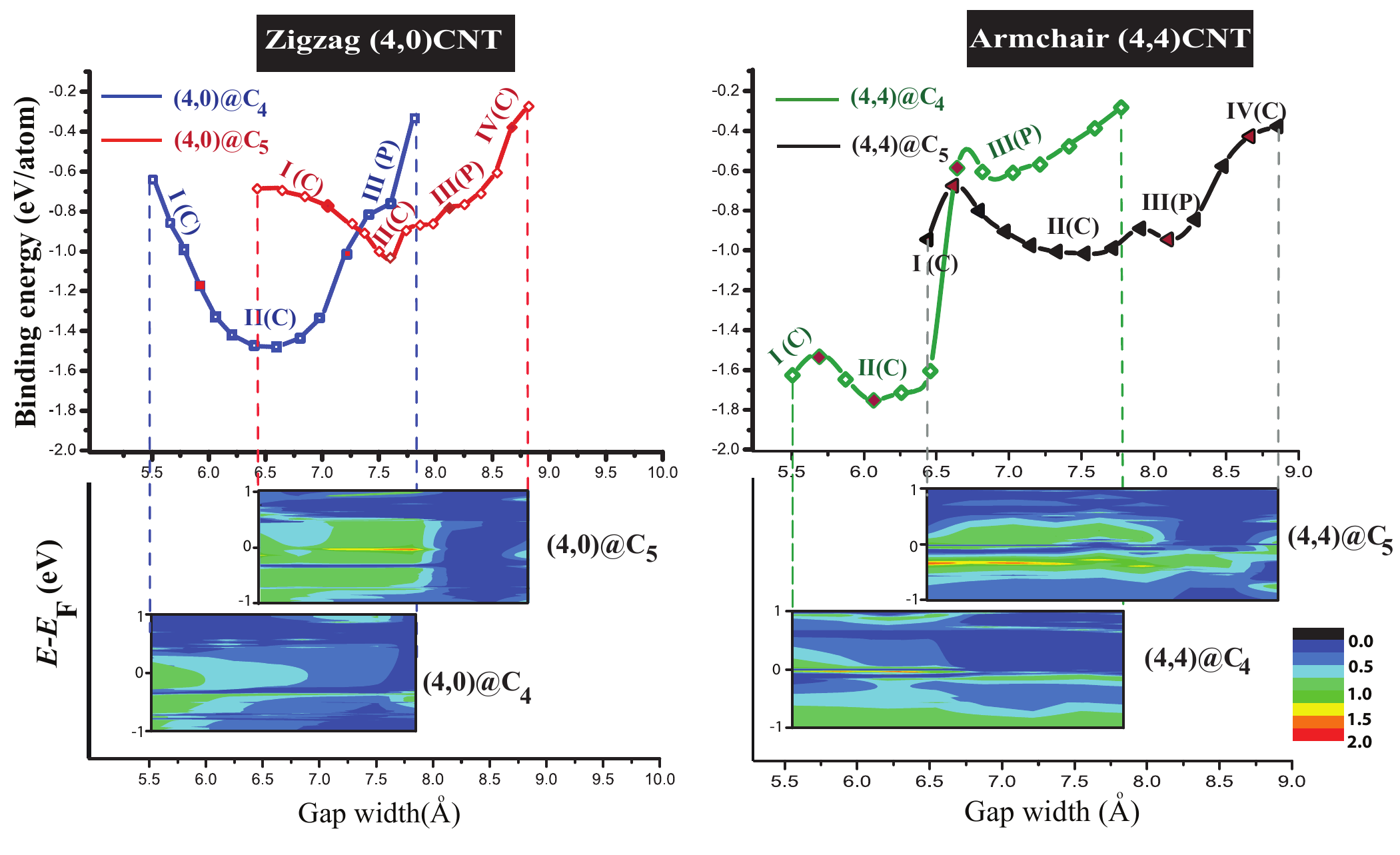}
\end{center}
\vspace{-20pt}
\caption{Binding energies vs gap width of the C$_4$ and C$_5$ wires connected to the (4,0)CNT (left panel) and (4,4)CNT (right panel) electrodes, and corresponding zero-bias transmission on the energy-length plane $T(E, L)$ (bottom). The location of a transition state of structures is marked by red dots (shown in the upper panel). The length-dependent transmission is correlated with the binding energies and wire structures.}\label{binding}
\end{figure*}

Fig.~\ref{binding} presents the binding energies and the zero-bias transmission $T(E, V=0)$ projected on the energy-length plane of the C$_{4}$ and C$_{5}$ wires connected to the zigzag (4,0)CNTs and armchair (4,4)CNTs. The length-dependent transmission is correlated with the binding energies and wire structures. We show that varyation of the gap width results in a substantial change in the wire structure and contact geometries which in turn affects transport properties of the junction.

Let us first discuss the zero-bias transmission $T(E,V=0)$ of the zigzag junctions in some details. We note that the effect of (4,0)CNT curvature plays an important role in the transport properties; therefore, the metallic behavior of the (4,0)CNT electrode has been observed. For the C$_5$ wire-zigzag junction, shown in Fig.~\ref{binding}a, the cumulene wires (4,0)@C$_{5}$-I(C) and (4,0)@C$_{5}$-II(C), show a broad resonance peak at the Fermi level, resulting in the high conductance. With stretching the junction, the wire changes structure to the polyyne, (4,0)@C$_{5}$-III(P), in which the intensity of the transmission peak drops to almost zero around the Fermi level. Before breakage, the wire changes structure back to the cumulene one, in which the transmission peak around the Fermi level appears again. The average conductance values of each structure: (4,0)@C$_{5}$-I(C), (4,0)@C$_{5}$-II(C), (4,0)@C$_{5}$-III(P) and (4,0)@C$_{5}$-IV(C) are 0.9G$_{0}$, 1G$_{0}$, 0.06G$_{0}$ and 0.37G$_{0}$, respectively. From the $T(E,V=0)$ discussed above, the electronic transport properties differ for the zigzag and armchair junctions. Obviously, charge transport for the polyyne wire is strongly suppressed in  the zigzag case, indicating that a switching behavior will be expected at low bias voltage. We emphasize that, due to the high curvature of the (4,0)CNT, the polyyne structure is defined with 1.28 \AA \ - 1.42 \AA \ bond alternation.

For the C$_4$ wire-zigzag junction, its transport properties are similar to those of the C$_5$ wire. We obtain the broadened resonance peak at the Fermi level for the cumulene; (4,0)@C${_4}$-I(C) and (4,0)CNT@C$_{4}$-II(C), but there is no resonance peak for the polyyne; (4,0)@C${_4}$-III(P). The average conductance of each structure; (4,0)@C$_{4}$-I(C), (4,0)@C$_{4}$-II(C) and (4,0)@C$_{4}$-III(P), is 0.9G$_{0}$, 0.4G$_{0}$ and 0.01G$_{0}$, respectively. Note that the conductance of (4,0)@C$_{4}$-I(C) (bent C$_4$ wire) is very high, because of a short separation between the leads of the C$_4$ compressed wire.

Next, we discuss the zero-bias transmission function $T(E,V=0)$ of the armchair junction, as shown in Fig.~\ref{binding}b.  We find that the structural change in the wire through varying the gap width affects the value of $T(E,V=0)$. Both C$_4$ and C$_5$ wires show the transmission peaks around the Fermi energy. Unlike zigzag junction, the resonance peaks, lying below the Fermi level ($E< E_F$) of the armchair junction, do not disappear from changing the wire structure from cumulene to polyyne, but the transmission drops by a factor of 2. In the armchair case, there is not enough difference between the conductance of cumulene and polyyne wires to show the switching characteristics useful for electronic applications.

\begin{figure}[bp]
\begin{minipage}[tp]{1.0\linewidth}
\begin{center}
\includegraphics[scale = 0.2]{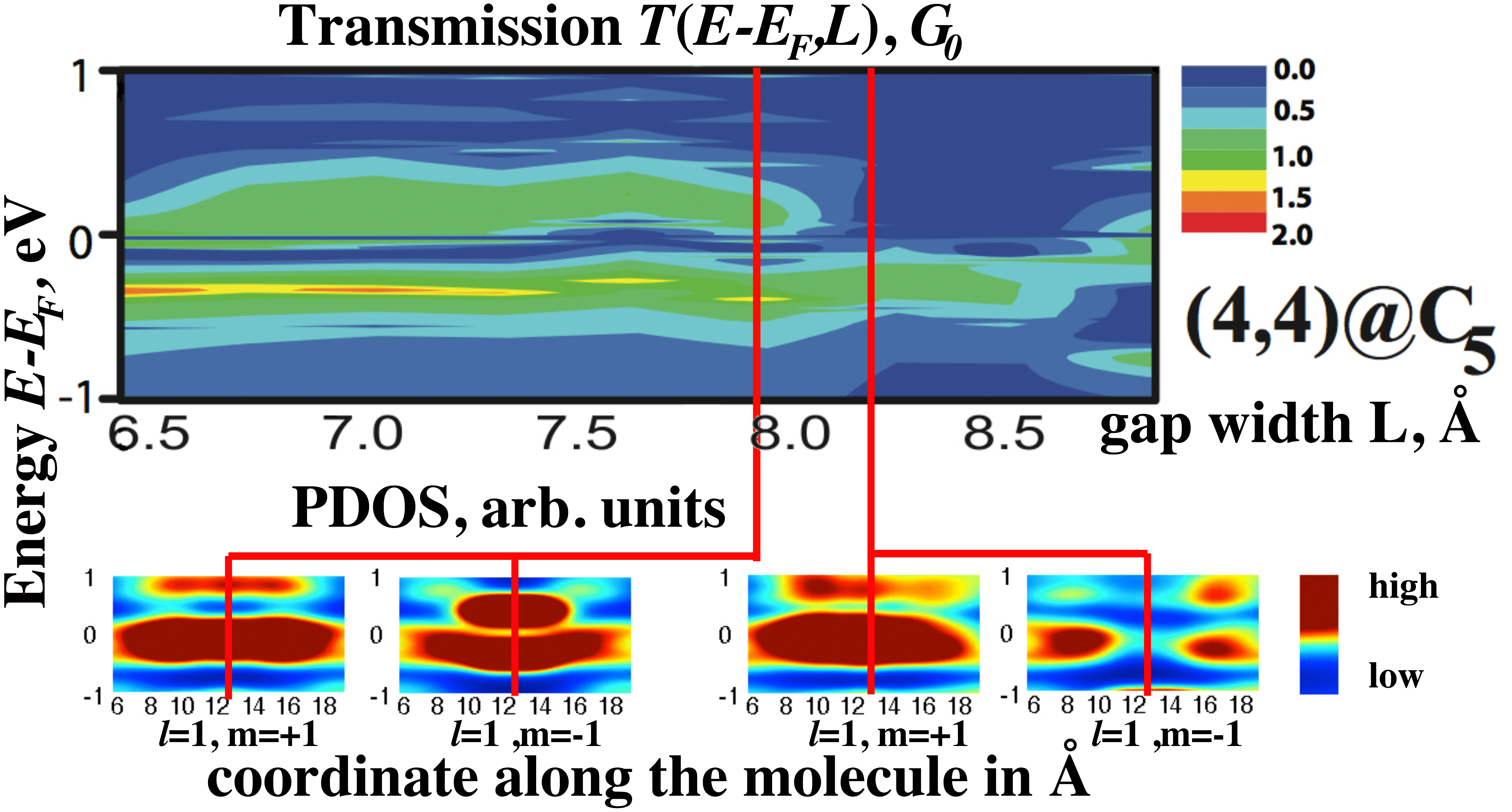}
\end{center}
\end{minipage}
\caption{Zero-bias transmission on the energy-length plane $T(E, L)$ and corresponding projected density of $p_{m=1}$ and $p_{m=-1}$ states for the carbon wire stretched between armchair (4,4)CNT electrodes at the two lengths corresponding to the II(C) and III(P) configurations as indicated in the Fig. \ref{binding}.}\label{tdc}
\end{figure}

The difference of the detailed transport behavior for both junctions is due to their entirely different electronic structures and bonding geometries. The bonding geometries are different for armchair and zigzag tubes (see Fig. \ref{sum}). It is seen that, for example, a tetragon is formed at the junction for the cumulene-zigzag junction; (4,0)@C$_{4,5}$-II(C), whereas the cumulene-armchair junction; (4,4)@C$_{4,5}$-II(C), has a pentagon. Also, the bonding geometries for the polyyne structure of both tubes are different.

\begin{figure}[tp]
\begin{center}
\includegraphics[width = 8 cm]{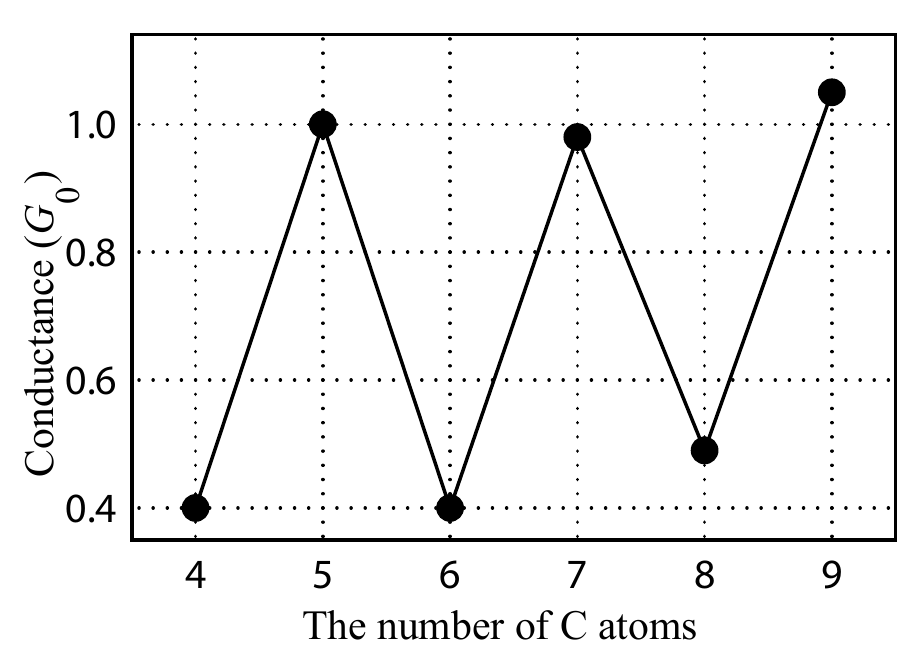}
\end{center}
\vspace{-20pt}
\caption{Zero-bias conductance of (4,0)@C$_{n}$-II, n= 4-9.}\label{oscillate}
\end{figure}

To elucidate the transport properties of the armchair junction, we show the $T(V=0,L)$ and corresponding projected density of the carbon wire stretched between armchair (4,4)CNTs (Fig.~\ref{tdc}). For the two different widths of the gap between the electrodes, the density of $p_{m=1}$ and $p_{m=-1}$ states along the molecule calculated with DFT method. The density of states correlates well with the $T(E)$ map since the carbon wire is directly coupled to the CNT. Molecular states with $p_{m=1}$ and $p_{m=-1}$ symmetry correspond to the two conducting channels in the shorter junction. In the longer junction, one of the C-C bonds breaks, and the density distribution changes for $p_{m=-1}$ states, effectively opening a wide gap in this channel and charging the molecule with $\approx$0.1e. We note that slight mismatch of the energies, the width of states and transmission resonances are due to the further adjusted charge of the system in NEGF calculation with open boundary conditions as compared to the neutral state of the system in DFT calculation of projected DOS.



The large difference in conductance between the cumulene and polyyne can be understood from the electronic structure of the junction: electrons tunnel through extended $p_x$ and $p_y$ orbitals delocalized both in the wire and in the CNT. In the case of armchair junction, the cumulene configuration shows the states contributing to both channels lying close to the Fermi level of the system. In the polyyne configuration, a gap over 2 eV opens in one of the channels with the $p$ orbitals lying along the CNT surface tangent plane, which reduces the number of conduction channels available close the Fermi level from two to one. In the case of the zigzag junction, the states remaining close to the Fermi level in polyyne configuration become additionally localized. Localization occurs when the defined single bond is created from last atom in the chain towards CNT. As we have discussed earlier, because of the CNT curvature, the wire connected to the tube becomes effectively capped. This leads to localization of the previously delocalized $p$-bonding along the CNT-carbon wire-CNT structure.

\begin{figure}[tp]
\begin{center}
\includegraphics[width = 8 cm]{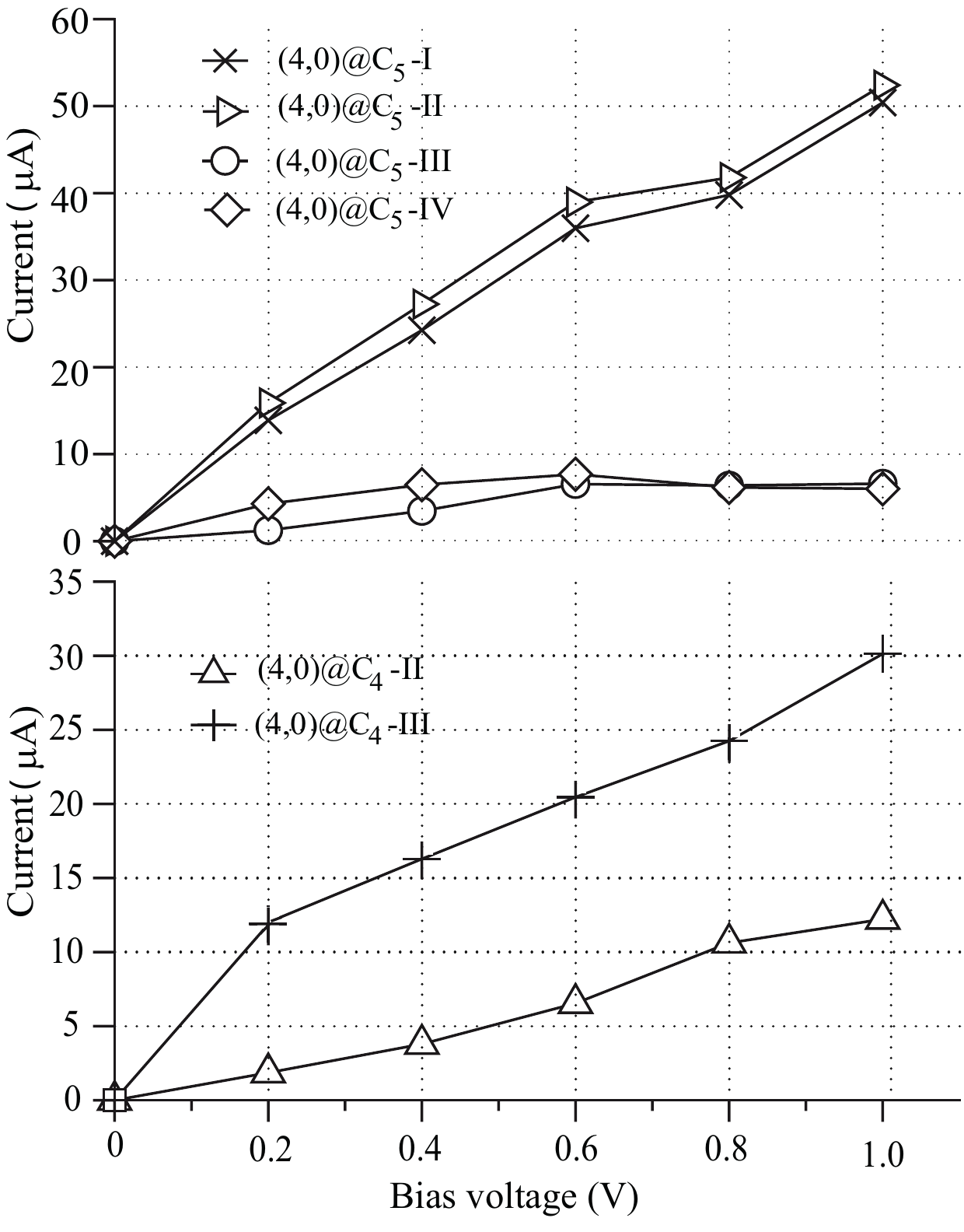}
\end{center}
\vspace{-20pt}
\caption{The $I-V$ characteristics of (a) (4,0)@C$_{4}$ (bottom panel) (b) (4,0)@C$_{5}$ (top panel) varying the gap width.}\label{iv}
\end{figure}
 
Fig.\ref{oscillate} exhibits the calculated zero-bias conductance of (4,0)@C$_{n}$-II(C), n= 4-9. Note that we selected (4,0)@C$_{n}$-II(C) in studying the oscillating conductance of the cumulene wire because that structure is the straight cumulene wire providing the highest conductance. Similar to the cumulene wire connected to metal leads\cite{Prasongkit:2010p780, Zhang13}, we observe the oscillatory characteristics in conductance of the cumulene wire-CNT junction, resulting from the difference in the electronic properties between even- and odd-cumulene wires \cite{Lang:1998,Prasongkit:2010p780}. We find that the conductance of the odd-wire is higher than that of even-wire by $\sim$ 60 \%. The conductance values do not show any pronounced dependence on the wire length when its length increased from four to nine atoms, resulting from the ballistic transport character of electrons through the short wire\cite{Prasongkit:2010p780}.


 \begin{figure}[tp]
 \begin{center}
 \includegraphics[width = 6 cm]{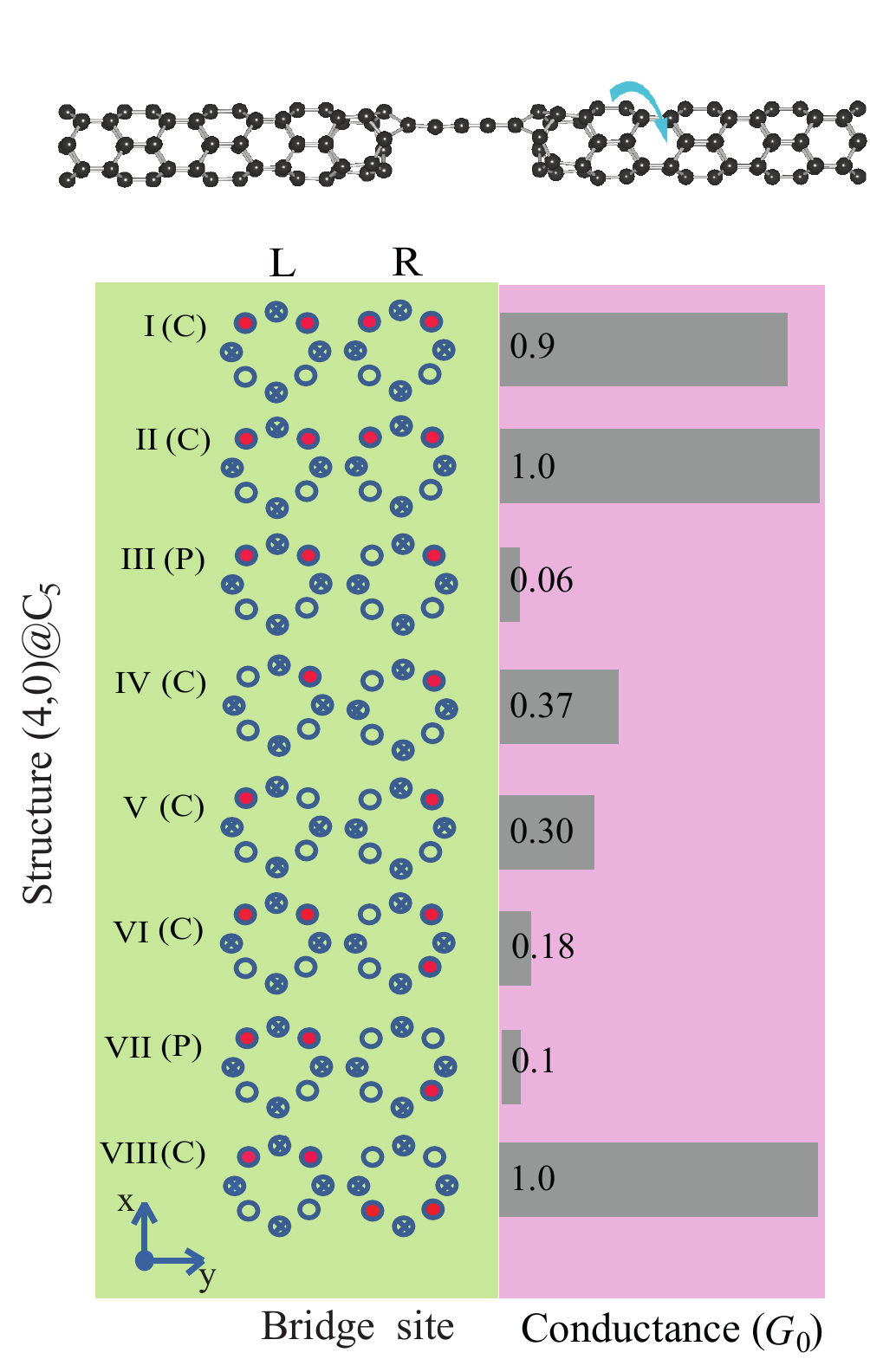}
 \end{center}
 \caption{The illustration of the C$_5$ wire bridging between (4,0)CNTs. By twisting the CNT, we can change bridging sites and bonding at the junction (top panel). The bridge sites of the left and right (4,0)CNT leads and the corresponding conductance of each structure (bottom panel).}\label{pdosb}
 \end{figure}

To compare our findings for small radii CNTs to the large ones, we note that the latter have low curvatures and are well represented by flat graphene electrodes. Graphene electrodes are also used in the recent TEM observations and electronic structure studies by Ref. \citenum{Jin:2009p592, Erdogan:2011bt} and others. We represent infinite radii of CNTs by studying the transport properties of the carbon wire connected to zigzag- and armchair-edge graphenes. For the zigzag graphene junction, we find that no zero-bias conductance is observed due to $\sim$ 2 eV band gap of the zigzag edge-graphene; thus, no current is expected at $V_b > 1$V. For the armchair graphene junction, the structural change of the carbon wires connected to the armchair edge graphene electrodes make the transmission resonance peak around the Fermi level shift in energy but with little change in its magnitude.

In the following, we will concentrate on a potential ON/OFF switch applications of the carbon wire-CNT junction. The electron transport properties of the carbon wire-(4,0)CNT junction will be discussed in the next section.

\subsection{(4,0)CNT-Carbon wire-(4,0)CNT}

To confirm the possibility of electrical switching behavior, as illustrated in Fig.\ref{iv}, we present the calculated $I-V$ characteristics (IVCs) of (a) (4,0)@C$_{5}$ (top panel) (b) (4,0)@C$_{4}$ (bottom panel), varying the gap width. We find that the ON/OFF current ratio of the (4,0)@C$_{5}$ is higher than that of the (4,0)@C$_{4}$. At $V_b$=0.2V, the ON/OFF ratio of (4,0)@C$_{4}$ and (4,0)@C$_{5}$ takes the value of $\sim$ 7 and 13, respectively, indicating electrical switching characteristics at low bias, coupled to the mechanical stretching of the CNT. We note that a very small diameter of the (4,0)CNTs plays an important role in their transport properties; thus, the metallic behavior of the (4,0)CNT electrode has been observed.

From the IVCs (Fig.\ref{iv}), the charge transport of (4,0)@C$_{4}$-III(P), (4,0)@C$_{5}$-III(P) and (4,0)@C$_{5}$-IV(C) is suppressed, whereas the (4,0)@C$_{4}$-II(C), (4,0)@C$_{5}$-I(C) and (4,0)@C$_{5}$-II(C) show a high-current state. It is important to notice at this point that the cumulene wires do not always give rise to the high conductance: a variation in conductance of the cumulene wire depends on the bonding geometries.

  \begin{figure}[tp]
   \begin{center}
   \includegraphics[width = 9 cm]{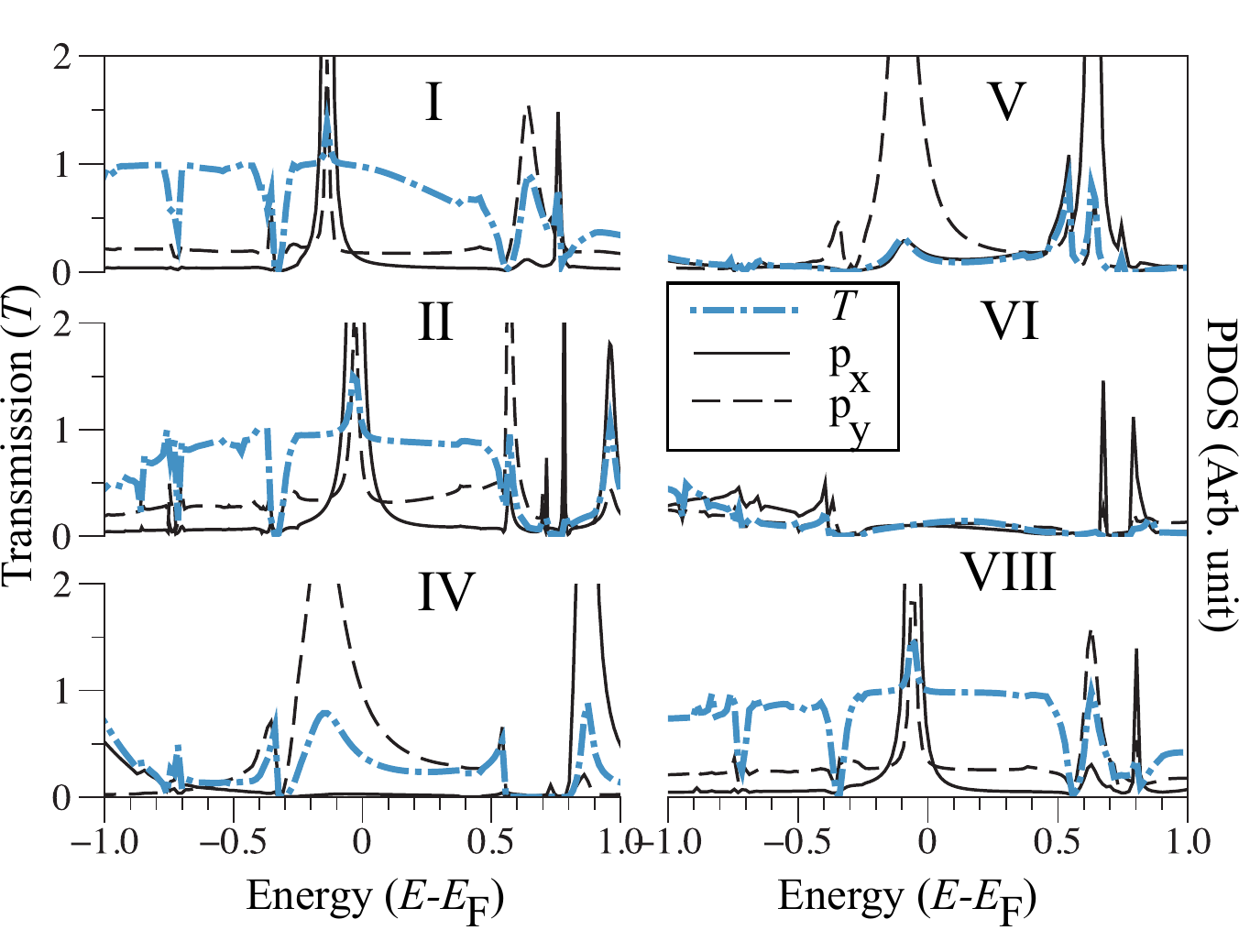}
   \end{center}
   \caption{The zero-bias transmission function and the PDOS of {$\mathrm{p_x}$} and {$\mathrm{p_y}$} orbitals of the cumulene C$_5$ wire, corresponding to the wire configuration in Fig. \ref{pdosb}.}\label{pdosa}
   \end{figure}

There is a probability that, in the experiment, the wire can jump to other sites when the CNT is effectively twisted. Consequently, we have investigated the geometrical structures and its corresponding conductance of the C$_5$ wire connected to all possible bridge sites on the (4,0)CNT, as presented in Fig. \ref{pdosb}. The wire structures have changed into either cumulenes or polyynes depending on the bridge site and gap width. Independently of the bridge site, we find that the polyyne wire is always a low conductance state. For the cumulene wire, interestingly, the results show a possibility to modulate the conductance of the zigzag junction by changing bridge sites, and bonding at the interface. The range of variation in conductance of the cumulene wires is $\simeq$ 0.1G$_0$-1G$_0$.

To explain the variation in conductance of the cumulene wire, we analyzed the transport channels via PDOS of the wire. We observe that the alignment of PDOS is determined by bridging sites and bonding at the junction. As we have already known that the cumulene wire has two $\pi$  orbitals around the Fermi level resulting from $\mathrm{p_x}$ and {$\mathrm{p_y}$} orbitals \cite{Shen2010,Prasongkit:2010p780}, the PDOS of the wire is decomposed into $\mathrm{p_x}$ and $\mathrm{p_y}$ components, as demonstrated in Fig. \ref{pdosa}. Our results can be classified into three cases according to bonding geometries at the interface. First, for the cumulene wire forming two bonds to each side of CNTs with two planes parallel to each other ((4,0)@C$_ 5$-I(C),II(C),VIII(C)), there are two states; $\mathrm{p_x}$ and  {$\mathrm{p_y}$}, with the same energy position at Fermi level. Consequently, electrons can propagate through $\mathrm{p_x}$ and {$\mathrm{p_y}$} eigenchannels of the carbon wire, showing a high conductance state. Apparently, the PDOS peak positions, consisting of the two transport channels, show the transmission function close to 2$G_0$. Second, for the cumulene forming one bond to the CNT leads; ((4,0)@C$_ 5$-IV(C),V(C)), there is the only {$\mathrm{p_y}$} state lying at the Fermi level, whereas the $\mathrm{p_x}$ state exists above the Fermi energy. This results in a decrease of conductance owing to only one transport channel at the Fermi level. Third, for the cumulene wire forming two bonds to the CNT lead with two planes perpendicular to each other; ((4,0)@C$_ 5$-VI(C)), the PDOS projected to $\mathrm{p_x}$ and  {$\mathrm{p_y}$} around the Fermi level is very low, causing a drop in conductance. We can therefore conclude that the variation in conductance of the cumulene wire is determined by the transport channels at the Fermi level, depending on the bridge sites and its bonding to the CNT leads.

\section{Summary}
We have performed first principles calculations to investigate the transport properties of the carbon wire between zigzag (4,0)CNTs and armchair (4,4)CNT electrodes. The gap width between the electrodes is varied and corresponding conductance variation upon the compression/elongation of the junction is calculated. Varying the gap width make the carbon wire change the structures (cumulene or polyyne) and contact geometries. We have observed the migration pathways of the carbon wire along the edge, which agrees well with the experimental results.

We find that the transport properties of the junction are significantly affected by the choice of chirality (zigzag or armchair). For the zigzag junction, the distinguishable ON- and OFF-state is observed, corresponding to the cumulene and polyyne structure, respectively. The zigzag CNT junction can be reversibly switched between ON- and OFF-state through varying the gap width between electrodes. In contrast, for the armchair junction, there is not enough difference in conductance to perform switching. The difference of the detailed transport behavior of both junctions is due to their entirely different electronic structures and bonding geometries, in which the high and low conductance states correspond to the cumulene and polyyne structures, respectively. In the studied configuration, the cumulene usually yields higher conductance than the polyyne wire. However, the cumulene wire can show the variation in conductance in the range from 0.1G$_0$ to 1G$_0$ upon the bridging geometries and sites. Their qualitative difference in transport mechanisms, resulting in the difference in the conductance state, was discussed by means of the transport channel at the Fermi level.

Oscillatory behavior in conductance of the cumulene wires with different lengths is demonstrated. We find that odd-cumulene wires yield higher conductance than the even-cumulene wires and ballistic transport behavior. The calculated ON/OFF ratios of the odd-wire are larger than that of the even-wire, resulting from the difference in the electronic properties between the odd- and even-wires.

Graphene leads, representing infinite radii of CNTs, have also been considered. However, graphene lead with zigzag edges has a band gap, and no zero-bias conductance is observed. Due to this semiconducting character of the zigzag electrode, switching behavior will be observed at a high bias voltage. In other aspects, the results for the graphene leads (infinite radii of CNT) are similar with that of CNT leads.

In contrast to metal electrodes, the CNTs can act as true nanoscale electrodes and there is a possibility that the carbon wire can jump to any site of the edge. We therefore investigate how the conductance is affected by the bonding site at the junction for the C$_5$ wire connected to (4,0)CNT electrodes. The range of variation in conductance of the cumulene wires is $\sim$ 0.1G$_0$--1G$_0$, revealing a potential to tune the conductance. The observed variation is explained by the PDOS analysis of p-orbital components of the cumulene wire. 

Finally, the carbon wire-zigzag CNT junctions demonstrate the prominent difference in conductance between ON and OFF states via compression/elongation junctions without breaking the wire and may be a promising candidate for mechano-switching device in molecular and nanoelectronics.

\begin{acknowledgments}
J.P. has partially been supported by the Nanotech- nology Center (NANOTEC), NSTDA, Ministry of Sci- ence and Technology, Thailand, through its program of Center of Excellence Network. A.G. and R.A. gratefully acknowledge financial support from Carl Tryggers Stiftelse f\"or Vetenskaplig Forskning and U3MEC, Uppsala. The calculations were performed at the high performance computing centers UPPMAX within the Swedish National Infrastructure for Computing.
\end{acknowledgments}

\bibliography{cntcum}

\end{document}